\documentclass[10pt,letterpaper]{article}
\usepackage{opex3}

\begin{document}

\title{Precise evaluation of polarization mode dispersion
by separation of even- and odd-order effects in quantum interferometry}

\author{A. Fraine$^{1}$, D.S. Simon$^{1,\ast}$, O. Minaeva$^{2}$, R. Egorov$^{1}$, A.V. Sergienko$^{1,3,4}$}

\address{$^1$ Dept. of Electrical and Computer Engineering, Boston
University, 8 Saint Mary's St., Boston, MA 02215.

$^2$ Dept. of Biomedical Engineering, Boston
University, 44 Cummington St., Boston, MA 02215.

$^3$ Photonics Center, Boston
University, 8 Saint Mary's St., Boston, MA 02215.

$^4$ Dept. of Physics, Boston University, 590 Commonwealth Ave., Boston, MA 02215.}

\email{simond@bu.edu}

\begin{abstract}
The use of quantum correlations between photons to separate measure even- and odd-order components of polarization mode dispersion (PMD) and chromatic dispersion in discrete optical elements is investigated. Two types of apparatus are discussed which use coincidence counting of entangled photon pairs to allow sub-femtosecond resolution for measurement of both PMD and chromatic dispersion. Group delays can be measured with a resolution of order 0.1 fs, whereas attosecond resolution can be achieved for phase delays.
\end{abstract}




\section{Introduction: Dispersion Measurement - Classical versus Quantum}

As optical communication networks migrate towards higher 40 Gbps and 100 Gbps data rates, system impairments due to dispersion, especially polarization mode dispersion (PMD), become a primary issue. This includes not only fiber PMD, but also contributions from switches, amplifiers, and all other components in the optical path. The fiber PMD and component PMD tend to accumulate in different manners as the size of the network grows.
In the long length regime, the differential group delay (DGD) due to fiber PMD has a known dependance on length, growing as $\sqrt{L}$ \cite{kogelnik}. In a similar manner, contributions from chromatic dispersion increase linearly in $L$. This known length dependence makes the dispersion of the optical fibers themselves relatively straightforward to measure and to take into account.

In contrast, component PMD was until recently considered to be too small in comparison to fiber PMD to affect significant penalties at the system level. Since the introduction of reconfigurable add-drop multiplexers (ROADMs), the number of components that could potentially contribute to the PMD in a given system has increased significantly.
Although the dispersive contribution of each separate component is relatively small, together they are capable of accumulating and of thereby making a significant contribution to the total system impairment. It is therefore important to be able to precisely and efficiently measure small values of DGD. However, since only fiber PMD was important in the past, no measuring techniques were developed for efficient evaluation of small DGD values. With component PMD starting to play a significant role, developing high-resolution evaluation of small PMD values in a single optical switch or other small discrete optical component represents a new challenge to optical researchers that must be addressed by modern optical metrology. In this paper, we address the measurement of dispersive effects in such discrete elements.

Polarization mode dispersion is the difference between wavenumbers of two orthogonal states of light at fixed wavelength, or equivalently, a polarization-dependent variation of a material's index of refraction. A number of methods have been developed for measuring it \cite{andresc,costa,poole2,poole,derickson,bakhshi,diddams,williams}.
Many traditional techniques for measuring PMD rely on an interferometric approach for high-resolution measurements of absolute values of optical delays. This approach requires one to use a monochromatic laser source and to keep track of the number of interference fringes. Therefore, the accuracy of the approach is limited by the stability of the interferometer, by the signal-to-noise level of the detector, and by the wavelength of the monochromatic radiation, leading to significant limitations. For example, the use of monochromatic classical polarized light does not allow one to measure the relative delay between two orthogonally polarized waves in a single measurement, so several measurements at different frequencies must be used to reconstruct the polarization dispersion properties of materials. The use of highly monochromatic laser sources creates the additional problem of multiple reflections and strong irregular interference that may have detrimental effect on measuring polarization dispersion.

White-light or low-coherence interferometry \cite{diddams} is another widely used approach. The ultimate resolution of such interferometric measurements will depend on the spectral bandwidth of the light source. Achieving sub-fs resolution in PMD measurement dictates the use of light sources with bandwidth in excess of 200 nm. Generating light of such a bandwidth with a smooth spectral profile is not an easy task in itself. Spectral modulations from existing sources with bumpy spectra produce 'ghost' features during measurement, leading to complications in dispersion evaluation. In addition, the visibility of interference with such super-broadband light is diminished due to dispersion effects.

Overall, while classical techniques can provide high-resolution measurement of polarization mode dispersion they still have limitations in many areas that quantum-based techniques can address.  For example, entangled photon states intrinsically provide an absolute value for polarization optical delay, in contrast to the conventional (classical) case, which is limited to determination of delay modulo an integer number of cycles of the light.
This is mainly due to the fact that quantum interferometry exploits both phase and group velocity effects in the same measurement \cite{branning,dauler}, a feat not possible in classical optics.

The current practical resolution of conventional dispersion evaluation techniques is limited to a few femtoseconds (fs).
The primary goal here is to use an interferometric setup with an entangled photon source to measure the component PMD of a small, discrete optical element to sub-femtosecond precision. Ideally, it would be desirable to measure chromatic dispersion with the same device, while allowing for the polarization and chromatic effects to be easily separable. We will show that this is indeed possible. Due to the frequency-anticorrelation in the entangled downconversion source used as illumination, we may independently determine the even-order and odd-order parts of the PMD's frequency dependence. Due to the reliance on the frequency anticorrelations within pairs of photons, the
separation method is intrinsically a two-photon quantum effect, and is not present in the  classical interferometer.

Classical attempts to simulate this even-odd separation effect by symmetrical chirping and anti-chirping of femtosecond laser pulses are constrained to a very narrow size of wavepacket thus making it not very practical.
The availability of such a separation is useful in a number of circumstances. One example is
when there is enough pulse broadening (second-order dispersion) to make accurate measurement of group velocity (first order dispersion) difficult. In a fiber, group velocity and broadening effects can be separated to some extent by simply taking a sufficiently long length of fiber as sample; the longer the fiber, the more accurately each can be measured. When dealing with switching elements or other small discrete optical elements, this option is not available. Another means must be found to prevent accurate measurement of the first-order group delay from being obscured by second-order broadening effects.
That is what is accomplished here: the location of a dip in the coincidence rate may be used to find the group velocity, and this location is unaffected by the second-order broadening as a result of the even-order dispersion cancellation.
Conversely, although the amount of broadening in a single small component may seem negligible, the total broadening from many such components present in a large network may be significant; thus high-accuracy measurements of these very small second-order dispersive contributions is important. Separating them off from the generally larger first-order contributions makes accurate measurements much easier.

We note that since the component being analyzed is assumed to be relatively small, the principal polarization axes may be assumed to
remain constant over the longitudinal length of the object and to be independent of frequency, with the dispersive contributions of the two polarization components remaining independent of each other. This greatly simplifies the analysis.

After a review of background and notation in section \ref{background}, three measurement
methods will be discussed in sections \ref{class}-\ref{typeb}.
The apparatus of section \ref{class} uses a single detector to
make a classical measurement; the system is illuminated with a broadband classical light source.
In contrast,  quantum measurements are made using two detectors connected in coincidence with illumination provided
by a source of entangled photon pairs (spontaneous parametric downconversion, (SPDC)). We will examine two quantum measurement
setups in sections \ref{typea} and \ref{typeb}.
In addition, in section \ref{typeb} we give a qualitative analysis that allows the positions of dips (or peaks) independently of the mathematical formalism.

The two quantum configurations will be distinguished from each other by referring to them as type A or type B.
They differ only in the presence or absence of a final beam splitter before detection, so they may both be
implemented in a single apparatus by allowing a beam splitter to be switched in or out of the optical path.
Similarly, by adding an additional polarizer and counting the singles rate at one detector instead of coincidence events, the classical
setup may also be implemented in the same device. Thus, a single apparatus could be made which is capable of performing any of the three types of measurements to be discussed.

This paper builds on two previous lines of work.
The apparatus used for the type A setup was introduced
previously \cite{branning,dauler}, where it was shown that quantum interferometry can achieve higher resolution than classical methods in measurements of PMD. Separately, the segregation of even- and odd-order chromatic dispersion effects was demonstrated in \cite{minaeva1}. Here, we bring the two strands together in a single device (type B), showing that we can separate even- and odd-order effects in PMD, as well as in chromatic dispersion, and that we can do so with the resolution available to the type A device.

As a further benefit of the quantum devices over classical methods, note that for the quantum cases there is no need to know in advance the principal axis directions of the device or object being measured.
Although the incoming photons are aligned along particular axes that are linked to a birefringent crystal orientation, their projections onto any rotated pair of orthogonal axes (including the principle axes of the sample) will remain equally entangled, allowing the method to work without any need to align the axes of the source and the device under test.

\section{Chromatic Dispersion and Polarization Mode Dispersion}\label{background}

First consider a material for which the index of refraction is independent of polarization. The frequency dependence of
the wavenumber $k={{2\pi n(\lambda )}\over \lambda}$
is given by a dispersion relation, which can be written near
some central frequency $\Omega_0$ as \begin{equation}k(\Omega_0\pm \omega )=k_0\pm \alpha\omega +\beta \omega^2 \pm \gamma\omega^3+\dots\label{chromatic}\end{equation} for $|\omega |<<\Omega_0$. The coefficients $\alpha$, $\beta ,\dots$ characterize the
{\it chromatic dispersion} or variation of the refractive index with frequency.
Explicitly, \begin{eqnarray} k_0&=&k(\Omega_0),
\qquad \qquad \qquad \alpha=\left.{{dk (\omega^\prime)}\over{d\omega^\prime }}\right|_{\omega^\prime =\Omega_0},\\ \beta&=&{1\over {2!}}\left.{{d^2k (\omega^\prime)}\over{d\omega^{\prime 2} }}\right|_{\omega^\prime =\Omega_0} ,\quad \qquad
\gamma={1\over {3!}}\left.{{d^3k (\omega^\prime)}\over{d\omega^{\prime 3} }}\right|_{\omega^\prime =\Omega_0} ,\; \dots \end{eqnarray}

Rather than looking at the individual terms in the expansion (\ref{chromatic}), we may also collect together all terms
containing even powers of $\omega$ and all terms containing odd powers to arrive at an expansion containing only two terms:
\begin{equation}k(\Omega_0+\omega )=k_{even}(\omega )+k_{odd}(\omega ),\end{equation} where
\begin{equation}k_{even}(\omega )=k_0+\beta \omega^2 +{\cal O}(\omega^4), \end{equation} and \begin{equation}  k_{odd}(\omega )=\alpha\omega +\gamma\omega^3+{\cal O}(\omega^5).
\end{equation}

In the case of nonzero {\it polarization mode dispersion} (PMD), the index of refraction varies with polarization. We now have two copies of the
dispersion relation, one for each independent polarization state:
\begin{eqnarray}k_H(\Omega_0\pm \omega )&=&k_{H0}\pm \alpha_H\omega +\beta_H \omega^2 +\dots \\ &=& k_{H,even}(\omega )+k_{H,odd}(\omega )\label{disprelH}\\
k_V(\Omega_0\pm \omega )&=&k_{V0}\pm \alpha_V\omega +\beta_V \omega^2 +\dots\\ &=& k_{V,even}(\omega )+k_{V,odd}(\omega ),\label{disprelV}\end{eqnarray}
where $H,V$ denote horizontal and vertical polarization.

To describe the PMD, we must define quantities that measure the differences between the two polarization states:
\begin{equation} \Delta k_0 =k_{V0}-k_{H0},\qquad \Delta \alpha =\alpha_V-\alpha_H, \qquad \Delta \beta =\beta_V-\beta_H .\label{PMDparam}\end{equation}
These parameters are defined per unit length. For the case of primary interest to us, discrete fixed-size objects, the formulas should really be written in terms of the relevant lumped quantities \begin{equation}\Delta \phi \equiv l\Delta k_0, \qquad
\Delta A \equiv l\Delta \alpha ,\qquad \Delta B \equiv l \Delta \beta ,\end{equation}
where $l$ is the axial thickness of the device under study.
However, we will continue to use the $\alpha$, $\beta$, and $\Delta k_0$ parameters of eq. \ref{PMDparam}, both because they are more commonly used, and because they allow easy comparison to the formulas used in fiber optics.

Note that $\Delta k_0={{\Omega_0\Delta n(\Omega_0)}\over c}$ is a measure of the difference in phase velocity between the two polarization modes, while $\Delta\alpha$ and $\Delta \beta$ are related to the difference in group velocity.
Also, it should be pointed out that the PMD and the chromatic dispersion are not entirely independent effects; in particular, the PMD coefficients themselves ($\Delta k_0,\; \Delta \alpha ,\; \Delta \beta$) are frequency dependent.

In the quantum cases, it is convenient to also define
$\tau_-=DL$, where $L$ is the thickness of the nonlinear downconversion crystal
and $D=u_0^{-1}-u_e^{-1}$ is the difference of the group velocities of the two
polarizations inside the crystal. We will restrict ourself to the simplest case of a bulk crystal, so the spectral distribution of the downconverted pairs is
described by the function \cite{rubin,klyshko} \begin{equation}\Phi (\omega )=\mbox{sinc}\left( {1\over 2}\tau_-\omega \right) ,\label{phidef}\end{equation} where the sinc function is defined by $\mbox{sinc}(x)={{\sin (x)}\over x}$. Photons are emitted from the downconversion process in frequency- anticorrelated pairs: the frequencies $\Omega_0\pm \omega$ in each pair are shifted equally, but in opposite directions, from the central frequency $\Omega_0=\omega_{pump}/2$, with the distribution of frequency shifts $\omega$ being given by $\Phi (\omega)$ of eq. \ref{phidef}. The downconversion time scale, $\tau_-$, is inversely proportional to
the spectral width of the source, and therefore determines the precision of the resulting measurements. The spectrum may be made wider by using a thinner nonlinear crystal, but this occurs at the expense of reducing the intensity of the downconverted light. High intensity and large bandwidth may be obtained simultaneously by use of a chirped crystal, although some of the details of the following analysis will then be changed.

\section{Classical PMD Measurement}\label{class}

\begin{figure}
\centering
\includegraphics[totalheight=2.0in]{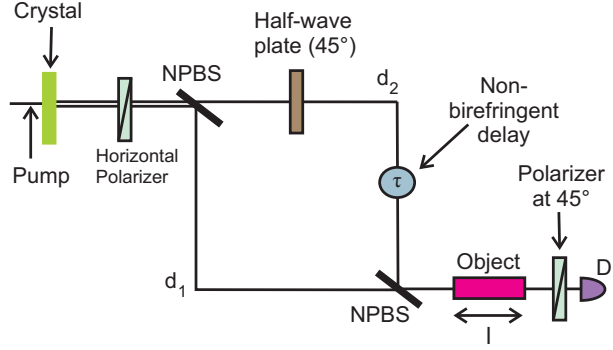}
\caption{\textit{Classical (single-detector) white-light setup for finding total PMD.}}\label{classicalfig}
\end{figure}

An apparatus equivalent to that shown schematically in fig. \ref{classicalfig} \cite{diddams} is commonly used to measure polarization mode dispersion. The illumination may be provided by any sufficiently broadband light source. For easier comparison with the later sections, we will assume the illumination is provided by type II parametric downconversion, but this is not necessary; since we use a single detector, the entanglement of the downconverted photons will play no role.

Assume an arbitrary amount of H and V polarization out of the downconversion crystal, so that the incident field in Jones vector notation is proportional to
\begin{equation}\int\left( \begin{array}{c} A_H (\omega )\\ A_V(\omega )\end{array}\right) \; d\omega  ,\end{equation} where $A_H$ and $A_V$ are the incoming amplitudes of the horizontal and vertical components.
After a horizontal polarizer, we destroy the quantum state and just pick off one component. We can think of it
as a classical broadband source of horizontally polarized light,  \begin{equation}\int\left( \begin{array}{c} A_H (\omega )\\ 0\end{array}\right) \; d\omega  .\end{equation}

For path 1 (lower), the horizontally polarized light accumulates a phase corresponding to the path length $d_1$.
For path 2 (upper), the horizontally polarized light passes through a $\lambda\over 2$ wave plate with fast axis $45^\circ$ from the horizontal, converting it into vertically polarized light,
\begin{equation}\left( \begin{array}{cc} 0 & -1 \\ 1& 0\end{array}\right) \int\left( \begin{array}{c} A_H (\omega )\\ 0\end{array}\right) \; d\omega =\int \left( \begin{array}{c} 0\\ A_H (\omega )\end{array}\right) \; d\omega  .\end{equation} In addition, the vertically polarized light in path 2 experiences a phase corresponding to the path length $d_2$ and an adjustable delay $\delta=c\tau_2$.

At the second beam splitter, the two components form a superposition of the form
\begin{equation}J_0= \int A_H(\omega ) \left( \begin{array}{c} e^{ik(\omega )d_1}\\e^{ik(\omega )(d_2+\delta )} \label{jnosample}\end{array}\right)\; d\omega ,\end{equation} with $k(\omega )={\omega \over c}$ (assuming the paths are in free space). In the absence of any sample after the second beam splitter,
this superposition will pass through a linear polarizer at $45^\circ$, resulting in
\begin{equation}J_0^\prime = \int {{A_H(\omega )}\over \sqrt{2}} \left( \begin{array}{c} e^{ik(\omega )d_1}+e^{ik(\omega )(d_2+\delta )}\\e^{ik(\omega )d_1}+e^{ik(\omega )(d_2+\delta )} \end{array}\right)\; d\omega ,\end{equation}

The intensity at the detector is then given by \begin{equation} I=|J_0^\prime |^2 =\int |A_H (\omega )|^2 \left[ 1+\cos \left( k(\omega )\left( \Delta d -\delta\right) \right)\right]\; d\omega .\end{equation} Here, $\Delta d=d_1-d_2$ is the path length difference between the two arms.

If a birefringent sample of length $l$ is introduced between the last beam splitter and the final polarizer, an additional polarization-dependent phase shift is added to the vector in eq. \ref{jnosample}:
\begin{equation}J_0= \int A_H(\omega ) \left( \begin{array}{c} e^{i\left[ k(\omega )d_1+k_H(\omega )l\right] }\\e^{\left[ ik(\omega )(d_2+\delta )+k_V(\omega )l\right]} \end{array}\right)\; d\omega .\end{equation} The resulting intensity at the detector is then:
\begin{equation} I=|J_0^\prime |^2  =\int |A_H (\omega )|^2 \left[ 1+\cos \left( {\omega \over c}\left( \Delta d -\delta\right) -\Delta k(\omega )l\right)\right]\; d\omega .\label{classintensity}\end{equation}

\begin{figure}
\centering
\includegraphics[totalheight=2.0in]{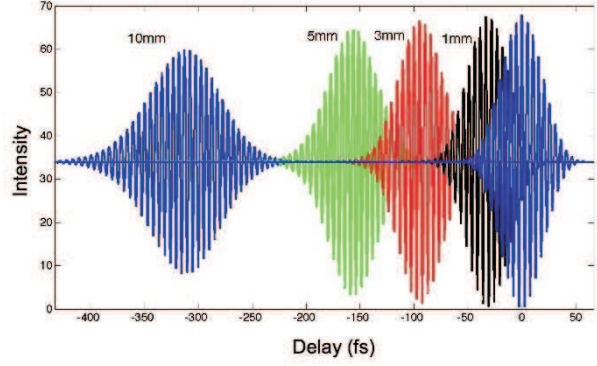}
\caption{\textit{Interferograms produced by apparatus of fig. \ref{classicalfig} for samples of different thicknesses. For a fixed thickness, the size of the shift may be used as a measure of the difference in phase velocities of the two polarizations.}}\label{classshiftfig}
\end{figure}

For Type II downconversion, the $A_H(\omega )$ and $A_V(\omega )$ are both proportional to $\Phi(\omega )=\mbox{sinc}\left( {1\over 2}\tau_-\omega \right)$. Plotting eq. \ref{classintensity} as a function of birefringent delay $\delta$ leads to interferograms such as those shown in fig. \ref{classshiftfig}.
Each interferogram will be phase shifted (moving the positions of the peaks and troughs {\it within} the envelope) due to the zeroth
order difference in dispersion $\Delta k_0$, while the envelope as a whole will be shifted horizontally due to the first order difference in dispersion $\Delta \alpha$ and broadened due to the second order difference $\Delta \beta$.  The interferograms shown in fig. \ref{classshiftfig} are shifted by different amounts due to the use of different sample thicknesses.
In this plot, a $200$ nm bandwidth centered at $1550$ nm was assumed, with a coherence length of $x_c={{\lambda_0^2}\over {\Delta \lambda}}=12\; \mu m.$



\section{Type A Quantum Measurement}\label{typea}

The goal now is to extract the polarization mode dispersion of an object with a higher precision than is possible with the classical apparatus of the previous section. In addition, we would like to be able to measure the even and odd orders of chromatic dispersion for each polarization.

\begin{figure}
\centering
\includegraphics[totalheight=2.0in, width=3in]{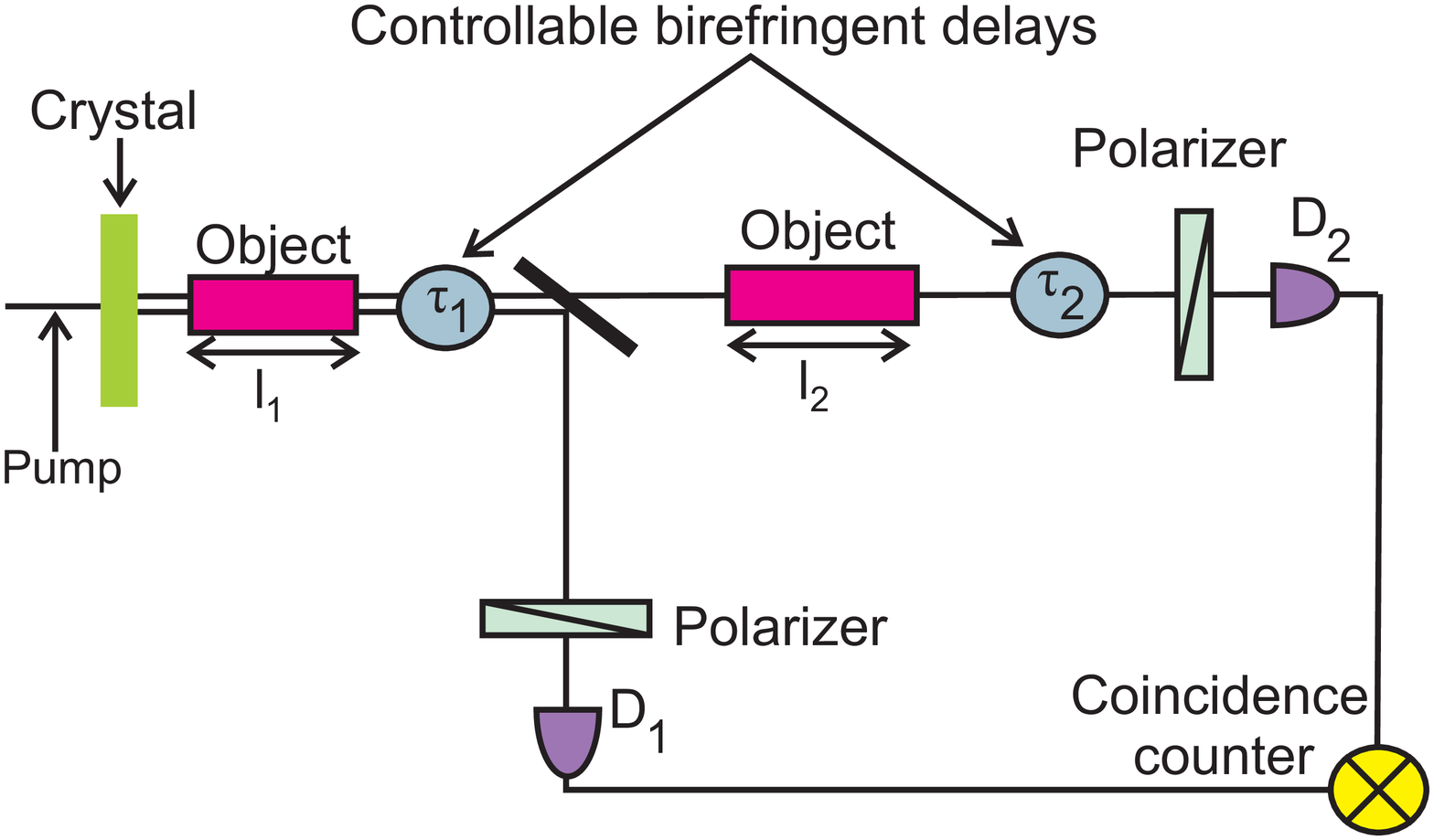}
\caption{\textit{Type A setup for measuring PMD parameters $\Delta \alpha\equiv \alpha_V-\alpha_H$ and $\Delta \beta \equiv \beta_V-\beta_H$.}}\label{pmdsetup1}
\end{figure}

The setup \cite{branning,dauler} is shown in fig. \ref{pmdsetup1}.
The downconversion is type II so that the
two photons have opposite polarization ($H$ and $V$). The photons have frequencies $\Omega_0\pm \omega$, where $2\Omega_0$ is the pump frequency.
Controllable birefringent time delays $\tau_1$ and $\tau_2$ are inserted before and after the beam splitter. Objects of lengths $l_1$ and $l_2$ may be placed before and after the beam splitter, respectively.  Polarizers at angles $\theta_1$ and $\theta_2$ from the horizontal are placed before the two detectors. In the following, we will take $\theta_1=\theta_2={\pi\over 4}$ and assume that the beam splitter is 50/50. Information about which polarization state travels in which branch of the apparatus will therefore be erased, allowing interference to occur with maximum visibility.

Rather than the Jones matrix formalism used in the previous section, it will be more convenient here to use creation and annihilation operators for horizontally and vertically polarized photons. The portion of the output from the downconversion process that is relevant to our purposes is the biphoton state \begin{equation}|\Psi \rangle =\int d\omega \; \Phi (\omega ) \hat a_H^\dagger (\Omega_0+\omega )\hat a_V^\dagger (\Omega_0-\omega )|0\rangle ,
\label{state}\end{equation} which will serve as the incident state of our setup. The positive-frequency parts of the fields at detectors $D_1$ and $D_2$, respectively, can be written in the forms
\begin{eqnarray} \hat E_1^{(+)} (t_1)&=& {1\over 2} \int d\omega \left\{ \hat a_H(\omega_1) e^{ik_H(\omega_1) l_1} + \hat a_V(\omega_1) e^{i[k_V(\omega_1) l_1+\omega_1\tau_1]}\right\} e^{-i\omega_1t_1}\\
\hat E_2^{(+)} (t_2)&=& {1\over 2} \int d\omega \left\{ \hat a_H(\omega_2) e^{ik_H(\omega_2) (l_1+l_2)} \right. \\ & &\qquad\qquad \left. + \hat a_V(\omega_2) e^{i[k_V(\omega_2) (l_1+l_2)
+\omega_2(\tau_1+\tau_2)]}\right\} e^{-i\omega_2(t_2+\tau )}.\nonumber
\end{eqnarray}
The coincidence rate is then computed by integrating the correlation function
\begin{equation} G^{(2)}(t_1,t_2) =\left| \langle 0| E_1^{(+)}(t_1) E_2^{(+)}(t_2)|\Psi \rangle |^2 \right|^2\end{equation}
over the characteristic time scale $T$ of the detectors: \begin{equation}R_c(\tau_1,\tau_2)=\int_{-T/2}^{T/2} dt_1dt_2 G^{(2)}(t_1,t_2).\label{rcfromG2}\end{equation} Since $T$ is generally much larger than the downconversion time $\tau_-$, we may safely simplify by taking $T\to \infty$.

Using eqs. \ref{state}-\ref{rcfromG2}, the coincidence rate may be written in the general form (\cite{rubin}) \begin{equation}R_c(\tau_1,\tau_2 )
=R_0\left\{ 1+C M(\tau_1,\tau_2)\right\},\label{rcgeneralform}\end{equation} where $R_0$ is a constant (delay-independent) background term and $C^{-1}=\int d\omega \left|\Phi (\omega )\right|^2={{2\pi}\over {\tau_-}}$. The dependence on the time delays is contained in
the modulation term \begin{eqnarray} M(\tau_1,\tau_2) &=& {1\over 2}\int d\omega
\left|\Phi (\omega )\right|^2 e^{-i\left[ \Delta k(\omega )-\Delta k(-\omega )\right] l_1-2i\omega \tau_1}\label{typeaamp}
\\ & & \qquad\qquad \times \; \left\{ e^{i\Delta k(-\omega )l_2 +i(\Omega_0-\omega )\tau_2}+
e^{-i\Delta k(\omega )l_2 -i(\Omega_0+\omega )\tau_2}\right\}\nonumber \\
&=&\int d\omega \left|\Phi (\omega )\right|^2 \\ & & \qquad\qquad \times \; \cos \left\{ \left[ \Delta k(\omega )-\Delta k(-\omega )\right] l_1 +\Delta k(\omega ) l_2 +2\omega\tau_1+(\Omega_0 +\omega )\tau_2 \right\}\nonumber
,\end{eqnarray} where the second form follows by changing the sign of the integration variable in the first term of the previous line. It can be seen that even-order PMD terms arising from the pre-beam splitter object cancel. Thus, measurements made with the object before the beam splitter will give us the odd-order PMD, and measurements made with the object after the beams splitter give the total PMD; making both measurements and then taking the difference will provide the even-order PMD. We can see the roles of the even and odd parts more clearly by splitting $\Delta k$ into its even and odd parts, then using the identity $\cos (A+B)=\cos A \cos B-\sin A \sin B$. The result is:


\begin{eqnarray} M (\tau_1,\tau_2 )  &=& \int d\omega \left|\Phi (\omega )\right|^2
\left\{ \cos \left[ \Delta k_{odd}(\omega ) (2l_1+l_2) +\omega (2\tau_1+\tau_2) \right]\cos \left[ \Delta k_{even}(\omega ) l_2 +\Omega_0\tau_2 \right] \right.\nonumber \\
& & \qquad -\sin \left. \left[ \Delta k_{odd}(\omega ) (2l_1+l_2)+\omega (\tau_1+\tau_2)\right]
\sin  \left[\Delta k_{even}(\omega ) l_2 +\Omega_0\tau_2 \right] \right\}.
\end{eqnarray}
Note that the integrand in the second term is odd in $\omega$, so the integral over that term vanishes. Therefore, this simplifies to
\begin{eqnarray} M (\tau_1,\tau_2 )  &=& \int d\omega \left|\Phi (\omega )\right|^2
\cos \left[ \Delta k_{odd}(\omega ) (2l_1+l_2)+\omega (2\tau_1+\tau_2) \right]
\nonumber \\
& & \qquad\qquad\qquad\qquad   \times \;\cos \left[ \Delta k_{even}(\omega ) l_2 +\Omega_0\tau_2  \right] . \end{eqnarray} We see that the even- and odd-order terms have separated into different cosine terms.

In the special case that $\Delta \beta$ and all higher order terms vanish, the integral of the
previous line can be done explicitly:
\begin{equation} M (\tau_1,\tau_2 ) = {{2\pi}\over {\tau_-}}
\cos \left[ \Delta k_0l_2+\Omega_0 \tau_2\right]
\Lambda \left[{{ \Delta \alpha (2l_1+l_2) +(2\tau_1+\tau_2)}\over {\tau_-}} \right] .
\end{equation}
In the last line we have used the result \begin{equation}\int d\omega \; \mbox{sinc}^2(a\omega )\cos\left(\omega\tau\right) ={\pi\over a}\Lambda\left( {\tau\over {2a}}\right),\end{equation} where
\begin {equation}
\Lambda (x) =
\bigg \{
\begin{array}{l}
1 - |x|, \qquad |x| \leq 1 \\
0, \qquad \qquad |x| > 1
\end{array}
\end{equation} is the unit triangle function.

\begin{figure}
\centering
\includegraphics[totalheight=2.0in, width=3in]{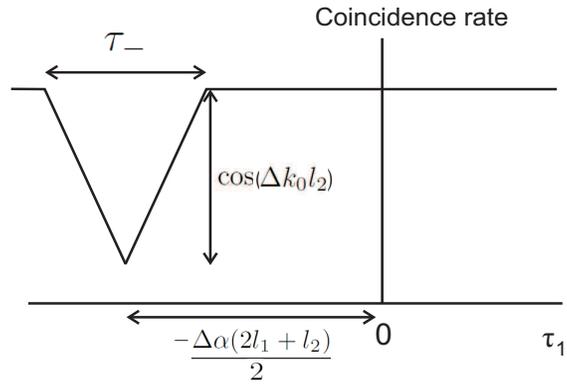}
\caption{\textit{Scanning over $\tau_1$ while keeping $\tau_2=0$. The horizontal shift of the minimum away from the origin determines $\Delta \alpha$, while the depth of the dip determines $\Delta k_0$.  The triangle function may lead either to a dip (as shown) or to a peak, depending on the sign of the cosine.}}\label{tau1scanfig}
\end{figure}

The coincidence rate is then
\begin{equation} R_c(\tau_1,\tau_2 ) =R_0\left\{ 1+\cos \left[ \Delta k_0l_2+\Omega_0 \tau_2\right]
\Lambda \left[{{ \Delta \alpha (2l_1+l_2) +(2\tau_1+\tau_2)}\over {\tau_-}} \right] \right\}\nonumber
.\label{lineara}\end{equation}  This result is
consistent with equation A31 of \cite{branning}, with the caveat that an extra
time delay $\tau_1$ has been added here.
We now have two possibilities: we can scan over $\tau_1$ while holding $\tau_2$ fixed, or vice-versa. If we scan over $\tau_1$ with $\tau_2=0$, we find a triangular dip similar to the HOM dip, as shown in fig. \ref{tau1scanfig}. The first order term in the PMD, $\Delta \alpha$ shifts the triangular envelope left or right, so that the bottom of the dip is at $\tau_1=-{{\Delta \alpha}\over 2}(2l_1+l_2)$; thus $\Delta \alpha$ may be determined by measuring the location of the minimum. The absolute value of the factor $
\cos \left( \Delta k_0l_2\right)$ in front of the triangle function gives the visibility of the dip;
so measuring the depth of the dip allows $\Delta k_0$ to be determined. Note that (depending on the sign of $\cos \left( \Delta k_0l_2\right)$) the "dip" may actually be a peak.

\begin{figure}
\centering
\includegraphics[totalheight=3.8in]{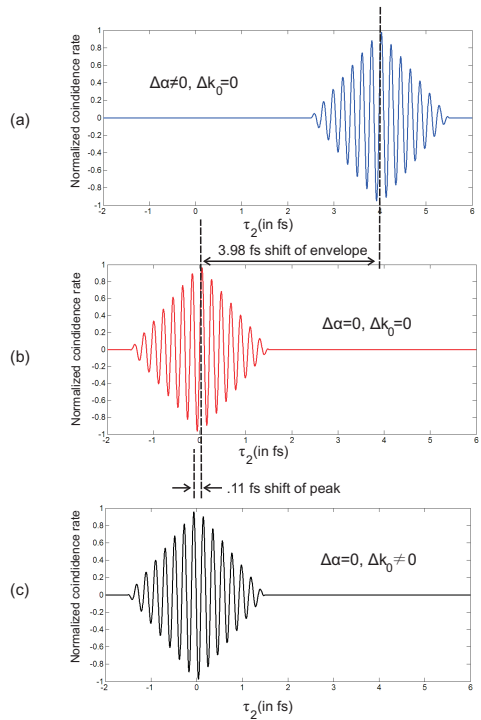}
\caption{\textit{Scanning over $\tau_2$ while keeping $\tau_1=0$. In (a), a nonzero $\Delta \alpha$ shifts the envelope from its position for $\Delta\alpha =0$ in part (b). The size of the shift can be measured with accuracy on the order of 0.1 fs. In part (c), a nonzero $\Delta k_0$
shifts the locations of the peaks within the unshifted envelope. The size of the shift can be measured with accuracy on the order of $.001\; fs\; =\; 1\; as$.}}\label{shiftingfig}
\end{figure}

Alternatively, we may scan over $\tau_2$ while holding $\tau_1=0$. This leads to an oscillating interference fringe pattern within the triangular envelope, similar to those of fig. \ref{classshiftfig}. The shift of the triangular envelope allows $\Delta \alpha$, the first order term in the PMD, to be determined as before. In this case, rather than determining visibility, the zeroth order term $\Delta k_0$ horizontally shifts the fringe pattern by distance $\tau_2={{\Delta k_0l_2}\over {\Omega_0}}$
within the envelope, allowing determination of $\Delta k_0$ from the size of this shift.
To see clearly the effects of each order of dispersion, fig. \ref{shiftingfig}
shows examples of such scans in the presence of zeroth-order and first-order dispersion separately. The fringes within the envelope as $\tau_2$ is scanned allow evaluation of the phase delays (the $\Delta k_0$ term) to an accuracy on the order of attoseconds ($10^{-18}$ s)  \cite{branning}. Group delays from the $\Delta \alpha$ term down to the order of 0.1 fs.

Note that only the {\it differences} of the horizontal and vertical polarization parameters ($\Delta \alpha$, $\Delta \beta$, etc.) appear in the formulas above.
The resulting interferogram is independent of the values of the parameters for fixed polarization $(\alpha_H$, $\alpha_V$, etc.) and so are insensitive to non-polarization-dependent dispersive effects.

In principle, Fourier transforming experimental data for the coincidence rate and then
fitting parameterized curves to it will allow the determination of higher order PMD parameters. However, this requires a large quantity of data to be
obtained at high precision. By
adding an additional beam splitter to the apparatus in the next section, we will arrive at a better method, which allows us to extract additional information; namely, it will also give us information about the $H$ and $V$ polarizations separately, not just their difference.

\section{Type B Quantum Measurement}\label{typeb}

The goal here is to see if additional information may be obtained with a variant of the previous
apparatus that mixes the final beams via an additional beam splitter.
This variation is inspired by the setup of ref. \cite{minaeva1}, in which even and odd portions of the
chromatic dispersion were separated
into different parts of an interferogram, allowing them to be studied independently of each other.

\begin{figure}
\centering
\includegraphics[totalheight=2.0in]{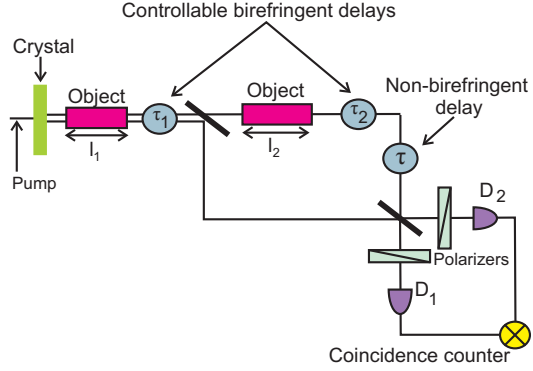}
\caption{\textit{Type B setup for finding even- and odd-order PMD}}\label{pmdsetup2}
\end{figure}

Consider the setup in fig. \ref{pmdsetup2}. This differs from the setup of the previous section (fig. \ref{pmdsetup1}) only by the addition of an extra beam splitter before the detectors and an additional nonbirefringent delay $\tau$ in one arm, after the first beam splitter. Two birefringent samples of lengths $l_1$ and $l_2$ are placed before and after the first beam
splitter. Birefringent delays $\tau_1$ and $\tau_2$ are present before and after the beam splitter as well, and a nonbirefringent
delay $\tau$ is added to one of the two arms after. For the sake of definiteness, assume that $\tau_1$ and $\tau_2$ delay the vertical (V) polarization and leave the horizontal (H) unaffected.
The system is illuminated with type II downconversion beams. The pump frequency is at $2\Omega_0$, while the signal and idler
frequencies will be written as $\Omega_0\pm \omega$. We will make use of the fact that the downconversion spectral function is
symmetric, \begin{equation}\Phi (\omega )=\Phi (-\omega).\label{phi}\end{equation} We will identify the $e$ and $o$ polarizations with $V$ and $H$ respectively.

It should be emphasized that in the notation used here, $\tau$ is an \emph{absolute }delay, so it must be positive. However, $\tau_1$ and $\tau_2$ are \emph{relative } delays of the vertically polarized photon compared to the horizontal, and so $\tau_1$ and $\tau_2$ may be positive or negative.

\begin{figure}
\centering
\includegraphics[totalheight=3.0in]{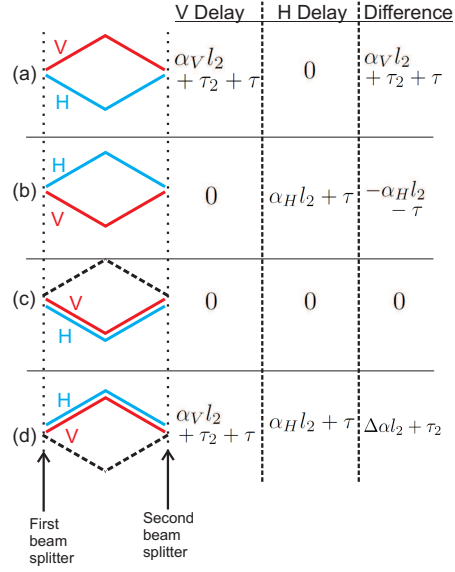}
\caption{\textit{The post-beam splitter delays corresponding to the four possible outcomes at the first beam splitter: one photon can go in each direction, with the vertical following the upper (a) or lower (b) path, both photons may follow the lower path (c), or both may follow the upper path (d). The second and third columns give the post-BS delays of the vertical and horizontal photons, respectively, while the final column gives the difference (vertical delay minus horizontal). }}\label{delaytable}
\end{figure}

We will find below that the effects of the even and odd orders separate and play different roles: the location of each dip in the interferogram (represented mathematically by a triangle function in the coincidence rate) is determined by the odd part, while the relative depths of the dips are controlled by the even part.
We may predict the number and location of each of these dips by identifying the ways in which it
becomes impossible from the relative timing of detection events in the two detectors to identify which photon took which path.
To do so, first note that the
delay between the $V$ and $H$ photons arising {\it before} the first beam splitter is \begin{equation}
\Delta \tau_{pre}\equiv \tau_V-\tau_H =\Delta \alpha\; l_1+\tau_1.\end{equation}
(This is the delay due to the object and $\tau_1$ alone; it is assumed that the intrinsic delay introduced by the known birefringence of the crystal itself has been compensated.)
There are four possible
ways in which the delay {\it after} the first beam splitter may compensate this pre-beam splitter delay, leaving a total delay of zero between the two
photons. These are enumerated in the table of figure \ref{delaytable}, which gives the total post-beam splitter delay $\Delta \tau_{post}$
for each case in the final
column. Setting \begin{equation}\Delta \tau_{pre}+\Delta \tau_{post}=0\end{equation} for these four possibilities predicts four dips in the coincidence rate at delay values for which the difference in the final column vanish; at these values, there is no path information available because the two photons arrive at the detector simultaneously, allowing for complete destructive interference between paths.

One additional dip (represented by the second triangle function in eq. \ref{linear} below) arises in a slightly different fashion. Here the time delay between the two photons is nonzero, but has a value that makes identification of the path impossible to identify. When the vertically-polarized photon V follows the lower path after the first beam splitter and the horizontally-polarized photon H takes the upper, the total relative delay between the two photons is \begin{equation}\Delta \tau_{total}\equiv \tau_V-\tau_H
=\Delta \alpha\; l_1+\tau_1=\Delta \tau_{pre}.\end{equation}
But when the two photons are interchanged (H along the lower path, V along the upper), the relative delay is  \begin{equation}\Delta \tau_{total}=\Delta \alpha\; (l_1+l_2)+
(\tau_1+\tau_2).\end{equation} If we require these to be negatives of each other (in other words, requiring $\Delta \tau_{post}=-2\Delta \tau_{pre}$) as in ref. \cite{branning}, we find the condition $\Delta \alpha\; (2l_1+l_2)+(2\tau_1+\tau_2)=0.$ This
leads to $\Delta \tau_{total}=-\Delta\tau_{pre}$. Because the photons arrive at different times and with different phases we see that in this case interference can occur, leading to the possibility that sines or cosines may modulate this term. These expectations will be explicitly verified below for the linearized case.

The coincidence rate may again be written in the general form of eq. \ref{rcgeneralform}. The delay-dependent modulation term will now be
\begin{eqnarray}& & M(\tau_1,\tau_2,\tau ) = \int d\omega \left|\Phi (\omega )\right|^2 e^{-2i\omega \tau_1}
e^{-2i\Delta k_{odd}(\omega )l_1}\label{r1exp}\\
&  &\times\; \left\{ 1- e^{-i \left[ \left( k_V\left(\Omega_0+\omega \right) -k_V\left(\Omega_0-\omega \right)\right)l_2
+2\omega (\tau +\tau_2)\right]}+e^{ i \left[ \left( k_H\left(\Omega_0+\omega \right) -k_V\left(\Omega_0+\omega \right)\right)l_2 -\left( \Omega_0+\omega\right) \tau_2\right]}\right. \nonumber \\ &  &  +e^{i \left[ \left( k_V\left(\Omega_0-\omega \right)-k_H\left(\Omega_0-\omega \right)\right)l_2 +\left( \Omega_0-\omega\right) \tau_2\right]} -e^{i \left[ \left( k_H\left(\Omega_0+\omega \right) -k_H\left(\Omega_0-\omega \right)\right)l_2 + 2\omega \tau \right]}\nonumber \\
& & -e^{i \left[ \left( k_H\left(\Omega_0+\omega \right) +k_V\left(\Omega_0-\omega \right)\right)l_2 +2 \Omega_0 \tau +\left(\Omega_0-\omega \right) \tau_2 \right]} -e^{-i \left[ \left( k_H\left(\Omega_0-\omega \right) +k_V\left(\Omega_0+\omega \right)\right)l_2 +2 \Omega_0 \tau +\left(\Omega_0+\omega \right) \tau_2 \right]}\nonumber\\
& & \left. +e^{i \left[ \left( k_H\left(\Omega_0+\omega \right) -k_H\left(\Omega_0-\omega \right)-k_V\left(\Omega_0+\omega \right)+k_V\left(\Omega_0-\omega \right)\right)l_2 -2\omega \tau_2 \right] }\right\} .
\nonumber\end{eqnarray}

\subsection{Linearized dispersion.}

To better understand eq. \ref{r1exp}, let's momentarily ignore the quadratic and higher order terms in the dispersion relations. Then we can write:

\begin{eqnarray} k_V(\Omega_0+\omega )&=&k_{V0}+\alpha_V\; \omega \\
k_H(\Omega_0+\omega )&=&k_{H0}+\alpha_H\; \omega   \\
\Delta k(\Omega_0+\omega )&=&\Delta k_0 +\Delta \alpha \; \omega
\end{eqnarray}

Even and odd parts become:
\vskip 5pt
\begin{center}\begin{tabular}{ccc} $k_{V,odd}=\alpha_V \; \omega$ & \qquad\qquad  & $k_{V,even}=k_{V0}$\\
$k_{H,odd}=\alpha_H \; \omega$ & \qquad & $k_{H,even}=k_{H0}$\\
$\Delta k_{odd} =\Delta \alpha \; \omega $ & \qquad & $\Delta k_{even}=\Delta k_0 $
\end{tabular}\end{center}
\vskip 5pt

This linear approximation allows all of the integrals of eq. \ref{r1exp} to be done analytically. For $\Phi (\omega )=\mbox{sinc}\left( {{\tau_-\omega}\over 2}\right) $, repeated use of the integral \begin{equation}\int d\omega \; \mbox{sinc}^2(a\omega )e^{i\omega (\tau +c)} ={\pi\over a}\Lambda\left(
{{\tau +c}\over {2a}}\right)\end{equation} leads to the result:

\begin{eqnarray}M(\tau_1,\tau_2,\tau )
& = &  {{2\pi}\over{\tau_-}} \left\{
\Lambda \left( {{2(\tau_1+\Delta \alpha l_1)}\over {\tau_-}}\right)\right. \label{linear}\\
& & +4\Lambda \left( {{2\tau_1+\Delta \alpha(2l_1+l_2)-\tau_2}\over {\tau_-}}\right)
\sin\left( k_{0V}l_2+\Omega_0(\tau+\tau_2)\right)\sin \left( k_{0H}l_2-\Omega_0\tau\right) \nonumber \\
& &  -\Lambda \left( {{2\left(\tau_1+\Delta \alpha l_1 +\alpha_Vl_2+\tau +\tau_2\right)}\over {\tau_-}}\right)
-\Lambda \left( {{2\left(\tau_1+\Delta \alpha l_1 -\alpha_Hl_2-\tau \right)}\over {\tau_-}}\right)
\nonumber \\ & &\left.
+\Lambda \left( {{2(\tau_1+\Delta \alpha (l_1+l_2) +\tau_2)}\over {\tau_-}}\right)
 \right\} .\nonumber  \end{eqnarray}
We see that we have five triangle functions centered at the locations predicted in the previous subsection; moreover, we see that one of these is modulated by sinesoidal functions as expected.

A number of specific methods may now be envisioned for extracting the chromatic and polarization mode dispersion parameters from this setup using various combinations of fixed and scanned delays. For example,
suppose that we scan over $\tau_1$, while holding $\tau$ and $\tau_2$ fixed. Then each of the $\Lambda $ factors above gives a triangular spike (of width $2\tau_-$) in the coincidence rate centered at the value of $\tau_1$ for which the argument of $\Lambda $ vanishes. We can then easily read off the locations of these spikes
from eq. \ref{linear}. Explicitly, the various terms of eq. \ref{linear} indicate that there should be triangular spikes centered at the values \begin{eqnarray}
\tau_1 &=& -\Delta \alpha l_1,\; \;  \;\; {1\over 2}\left[ \tau_2-{\Delta \alpha \left(2l_1+l_2\right) }\right] ,\; \;\;\;  -\left( \alpha_Vl_2+\Delta \alpha l_1 +\tau+\tau_2\right), \\
& & \qquad  \alpha_Hl_2+\tau -\Delta \alpha l_1 ,\;\;\;
\; \tau_2-\Delta \alpha (l_1+l_2) .\nonumber
\end{eqnarray}
So suppose we have a sample only after the beam splitter ($l_1=0$) and then we do three scans over $\tau_1$, each with different values of $\tau$ and $\tau_2$:

(i) Take $|\tau_2 |$ large, with $\tau=0$. Then the $\tau_2$-dependent peaks move far from the origin, off the edge of the plot.
We will be left with peaks at $\tau_1=0$ and at $\tau_1=\alpha_H l_2$; from the locations of the latter we can read off $\alpha_H $.

(ii) Take $|\tau_2 |$ and $|\tau |$ both large, but with$|\tau_2 +\tau |$ small. Then we will be left with peaks at $\tau_1=0$ and $\tau_1=
\alpha_V l_2$, so we can read off $\alpha_V $.

(iii) Take $|\tau |$ large, with $\tau_2=0$. We will be left with peaks at $\tau_1=0,$ and $
\Delta \alpha l_2$, so we can read off $\Delta \alpha $.

Since it is more difficult to achieve large values for the birefringent delays than for the nonbirefringent ones, this procedure may not always be the most practical. An alternative version will be described below when we examine a special case.

Finally, notice that some of the triangular spikes will have their heights modulated by cosine terms. The arguments of the cosines depend on $k_0$, so that measuring the heights of these spikes relative to the others
will allow $\Delta k_0$ to be determined as well.

\subsection{Adding in quadratic dispersion}

\begin{figure}
\centering
\includegraphics[totalheight=1.6in]{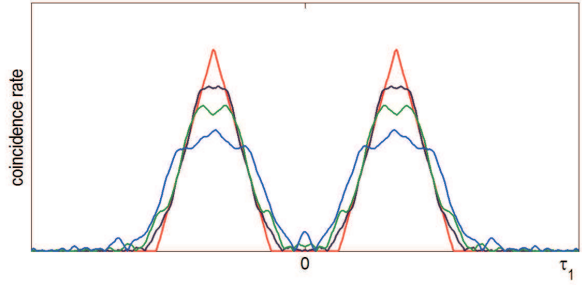}
\caption{\textit{Effect of quadratic dispersion term $\Delta\beta$ on a pair of triangular peaks.
The red curve is for $\Delta\beta=0$, the lower curves correspond to increasing values of $\Delta \beta$ for fixed $\Delta\alpha$ and $\Delta k_0$. }}\label{evenfig}
\end{figure}

When the quadratic ($\Delta \beta$) term is added back in, analytic expressions can no longer be obtained and numerical simulations must be done.
An example is shown for one pair of triangular peaks in fig. \ref{evenfig}.
In the figure, unrealistically large values of $\Delta \beta$ were used to make the effect clearly visible. For $\Delta\beta=0$ (red curve),
the peaks have the same triangular form predicted earlier. As $\Delta \beta$ increased for fixed $\Delta\alpha$ and $\Delta k_0$ the top of the
triangle flattens and gains small oscillatory features; the triangle also broadens slightly.

For realistic values of $\Delta k_0$ and $\Delta \beta$, the alteration of the peak's height by $\Delta \beta$ is negligible, so that the height of the
peak can still be used to measure $\Delta k_0$. The most straightforward method to separate the value of $\Delta \beta$ from $\Delta k_0$ is to fit a parameterized curve to the data and look for the values of the parameters $\Delta k_0$ and $\Delta \beta$ that give the best fit.

\subsection{Example: Postponed delay only}

As a special case, we can look at the situation where there is no sample or delay before the first beam splitter. This is accomplished by setting $l_1=\tau_1=0$. (In reality, the downconversion crystal itself acts as a birefringent sample before the beam splitter, but a fixed $\tau_1$ may be inserted to cancel it, so that without loss of generality, we can still take the combination $\Delta \alpha_{crystal}l_1+\tau_1=0$.)

\begin{figure}
\centering
\includegraphics[totalheight=2.0in]{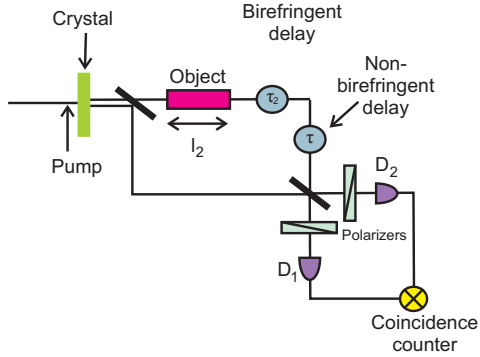}
\caption{\textit{Setup with object and delays only after the first beam splitter. }}\label{afterfig}
\end{figure}

The setup now looks as shown in fig. \ref{afterfig}, and the coincidence rate as given by eqs. \ref{rcgeneralform} and \ref{linear} is:
\begin{eqnarray}& & R(\tau ,\tau_2) = R_0\left\{ 2 +4\Lambda \left( {{\Delta\alpha l_2-\tau_2}\over {\tau_-}}\right) \sin\left[
k_{0V}l_2+\Omega_0(\tau+\tau_2)\right] \sin\left[ k_{0H}l_2-\Omega_0\tau \right] \right.\nonumber \\
& & \qquad \left. -\Lambda \left( {{2(\alpha_V l_2+\tau +\tau_2)}\over {\tau_-}}\right) -\Lambda \left( {{2(\alpha_H l_2+\tau )}\over {\tau_-}}\right)  +\Lambda \left( {{\Delta\alpha l_2+\tau_2}\over {\tau_-}}\right) .\right\}\nonumber
\end{eqnarray}

\begin{figure}
\centering
\includegraphics[totalheight=2.0in]{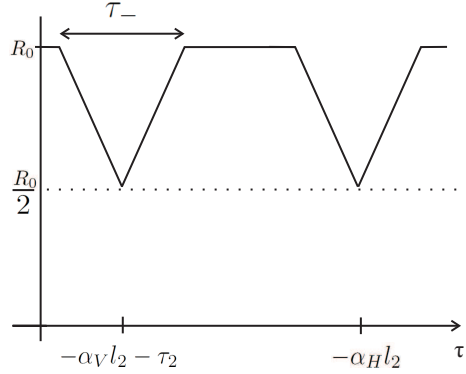}
\caption{\textit{Scan over $\tau$ for fixed $\tau_2$, when there is an object only after the first beam splitter.}}\label{tauscan1fig}
\end{figure}

Holding the remaining birefringent delay $\tau_2$ fixed and scanning over the nonbirefringent delay $\tau$,
there should be dips at \begin{equation}\tau =-\alpha_Vl_2-\tau_2 \quad \mbox{ and } \quad \tau=-\alpha_Hl_2 ,\end{equation}
as in fig. \ref{tauscan1fig}.
So, running two scans over $\tau$ using two different values of $\tau_2$, the location of the dip that remains at the same position in both scans gives us the value of $\alpha_H$. The other dip moves between the scans; measuring its location during either scan will then give the value of $\alpha_V$. $\Delta \alpha $ is then given by the difference between the two measured values.

If it is possible to arrange a value of $\tau_2$ large enough to satisfy $\tau_2=\Delta \alpha l_2$, then the term with the sines will be large, in which case we may also be able to extract $k_{0H}$ and $k_{0V}$ by scanning $\tau$ over a range of values for which the other triangle functions vanish and fitting the resulting data curve to the function $\sin\left[
k_{0V}l_2+\Omega_0(\tau+\Delta \alpha l_2)\right] \sin\left[ k_{0H}l_2-\Omega_0\tau )\right]$. Alternatively, if
only $\Delta k_0$ is needed (not $k_{0H}$ and $k_{0V}$ separately), it may be simpler to
remove the final beam splitter (turning the type B apparatus back into type A), then scanning over $\tau_2$ and find $\Delta k_0$ from the shift in oscillation fringes via equation eq. \ref{lineara}.

\section{Conclusions}
We have demonstrated that it is possible to separately measure the even and odd-order contributions to the chromatic dispersion and the polarization mode dispersion, and that such measurement have higher accuracy than dispersion measurements accomplished with classical white-light interferometry or time-of-flight methods. Summarizing once again the available resolutions of the classical and quantum methods, the resolution of group delay measurements are limited by (i) the ability to localize the minimum of the envelope in the classical method (resolution of a few fs in the plots of fig. 2), and (ii) the downconversion time scale $\tau_-$ in the quantum case (typically tenths of a femtosecond).

The Type B quantum method provides the high resolution previously available in the type A method, but with additional advantage of being able to separate even- and odd-order effects for both PMD and chromatic dispersion, without the need for aligning the principal axes of the device with the polarization directions of the source photons.
In each case, though, separating the various orders within the even-order part (separating zeroth order from second order, for example) or within the odd-order part is a much more difficult problem, which has yet to be solved in a fully satisfying manner. As a final point, note that both the type A and type B devices are truly quantum, in that a frequency-entangled photon source is required in order for them to operate.
If a classical source is used, then the frequencies in the two branches will not appear in anticorrelated pairs $\Omega_0+\omega$
and $\Omega -\omega$ (for example in eqs. \ref{typeaamp} and \ref{r1exp}), so that the required cancellations of even or odd orders will not occur in the various terms.

\section*{Acknowledgements}
This research was supported by a grant from Capella Intelligent
Systems, Inc., and by the DARPA InPho program through US
Army Research Office award W911NF-10-1-0404.

\end{document}